\documentclass[12pt]{article}
\usepackage{graphicx}
\usepackage{amsmath}
\input psfig.sty

\begin{document}

\begin{titlepage}

\title{\bf Accelerating universe as from \\
the evolution of extra dimensions\\
\vspace{0.2cm}}

\author{Je-An \ Gu\thanks{%
E-mail address: jagu@phys.ntu.edu.tw} \ \ \ and \ \ W-Y. P. Hwang\thanks{%
E-mail address: wyhwang@phys.ntu.edu.tw} \\
{\small Department of Physics, National Taiwan University, Taipei
106, Taiwan, R.O.C.}
\medskip
}

\date{\small September 24, 2002}

\maketitle

\begin{abstract}
In this paper we propose that the accelerating expansion of the
present matter-dominated universe, as suggested by the recent
distance measurements of type Ia supernovae, is generated along
with the evolution of space in extra dimensions. The Einstein
equations are first analyzed qualitatively and then solved
numerically, so as to exhibit explicitly these patterns of the
accelerating expansion in this scenario. A fine-tuning problem
associated with such a scenario is also described and discussed.
\end{abstract}


\end{titlepage}

\section{Introduction}

The recent distance measurements of type Ia supernovae suggest an
accelerating expansion of the present universe
\cite{Perlmutter:1999np,Riess:1998cb}. In many of the current
cosmological models, the present accelerating expansion is driven
by an energy source called ``dark energy'', with a positive
cosmological constant \cite{Lambda models} or the so-called
``quintessence'' (a slowly evolving scalar field
\cite{Caldwell:1998ii,ComplexQ}) as a possible candidate. Instead
of attributing this acceleration to the mysterious dark energy, we
consider in this paper the possible existence of extra spatial
dimensions and explore the feasibility of producing the present
accelerating expansion via the evolution of these extra
dimensions.

The application of extra dimensions is a general feature in
theories beyond the standard model, especially in theories for
unifying gravity and other forces, such as superstring theory.
These extra dimensions should be ``hidden'' for consistency with
observations. Various scenarios for ``hidden'' extra dimensions
have been proposed, for example, a brane world with large compact
extra dimensions in factorizable geometry proposed by Arkani-Hamed
\emph{et al.} \cite{Arkani-Hamed} (see also
\cite{Antoniadis:1990ew}), and a brane world with noncompact extra
dimensions in nonfactorizable geometry proposed by Randall and
Sundrum \cite{Randall&Sundrum}. In this paper, we employ the
simplest scenario: small compact extra dimensions in factorizable
geometry, as introduced in the Kaluza-Klein theories
\cite{Kaluza&Klein}.

We study spatially homogeneous, isotropic, perfect-fluid
cosmological models in $(1+3+n)$ dimensions where $n$ is the
number of extra dimensions. In Sec.\ \ref{general features}, we
first obtain, from the Einstein equations, some general features
of the evolution of the higher-dimensional space, especially for a
radiation-dominated universe and a (nonrelativistic)
matter-dominated universe. In Sec.\ \ref{accel 3-space}, we then
explore the possibility of producing an accelerating expansion of
the ordinary three-space via the evolution of `extra space' for a
matter-dominated universe. We analyze the Einstein equations to
show qualitatively the behavior of this evolution and obtain
numerical solutions which illustrate explicitly the accelerating
expansion of the ordinary three-space along with the collapse of
the extra space. We note that, while the Kaluza-Klein cosmology
and inflation in higher dimensional space-time in connection with
the early universe were discussed widely in the 1980s
\cite{Freund:1982pg,KK-cosmology} (for a review, see
\cite{Kolb:1990vq}; for recent work, see
\cite{Arkani-Hamed:1999gq,Mongan:2001cr}), the focus of this paper
has to do with the present accelerating matter-dominated universe.

\section{Evolution of Spaces in Ordinary Three Dimensions and Extra Dimensions: General Features\label{general features}}


We consider a metric of a (3+n+1)-dimensional space-time, in which
both the ordinary three-space and the extra (or ``hidden'') space
are homogeneous and isotropic:
\begin{equation}
ds^2 = dt^2 - a^2(t) \left( \frac{dr_a^2}{1-k_a r_a^2} + r_a^2 d
\Omega_a^2 \right) - b^2(t) \left( \frac{dr_b^2}{1-k_b r_b^2} +
r_b^2 d \Omega_b^2 \right) , \; 0 \leq r_b \leq 1  %
\label{metric with extra dim}
\end{equation}
where $a(t)$ and $b(t)$ are scale factors, and the values of $k_a$
and $k_b$ are related to the curvatures of the ordinary
three-space and the extra space, respectively. Assuming that the
matter content in this higher-dimensional space is taken to be a
perfect fluid, we can write the Einstein equations, which govern
the evolution of the ordinary three-space and the extra space, as
\begin{equation}
3 \left[ \left( \frac{\dot{a}}{a}\right)^2 + \frac{k_a}{a^2}
\right] + \frac{n(n-1)}{2} \left[ \left(
\frac{\dot{b}}{b}\right)^2 + \frac{k_b}{b^2} \right] +
3n\frac{\dot{a}}{a} \frac{\dot{b}}{b} = 8 \pi \bar{G} \bar{\rho}
\, , \label{G00 eq with extra dim}
\end{equation}
\begin{equation}
2 \frac{\ddot{a}}{a} + n\frac{\ddot{b}}{b} + \left[ \left(
\frac{\dot{a}}{a} \right)^2 + \frac{k_a}{a^2} \right] +
\frac{n(n-1)}{2} \left[ \left( \frac{\dot{b}}{b}\right)^2 +
\frac{k_b}{b^2} \right] + 2n \left( \frac{\dot{a}}{a} \right)
\left( \frac{\dot{b}}{b} \right) = - 8 \pi \bar{G} \bar{p}_a \, ,
\label{Gii eq with extra dim}
\end{equation}
\begin{eqnarray}
3 \frac{\ddot{a}}{a} + (n-1)\frac{\ddot{b}}{b} + 3 \left[ \left(
\frac{\dot{a}}{a} \right)^2 + \frac{k_a}{a^2} \right] +
\frac{(n-1)(n-2)}{2} \left[ \left( \frac{\dot{b}}{b}\right)^2 +
\frac{k_b}{b^2} \right] && \nonumber \\
+ 3(n-1)\left( \frac{\dot{a}}{a} \right)
\left( \frac{\dot{b}}{b} \right) = - 8 \pi \bar{G} \bar{p}_b \, , && %
\label{Gjj eq with extra dim}
\end{eqnarray}
where $\bar{G}$ and $\bar{\rho}$ are the gravitational constant
and the energy density in the higher-dimensional world, and
$\bar{p}_a$ and $\bar{p}_b$ are the pressures in the ordinary
three-space and the extra space, respectively. Assuming simple
equations of state $\bar{p}_a=\omega_a \bar{\rho}_a$ and
$\bar{p}_b=\omega_b \bar{\rho}_b$ with constant state parameters
$\omega_a$ and $\omega_b$, the conservation of stress energy gives
rise to
\begin{equation}
\bar{\rho} \propto a^{-3(1+\omega_a)} b^{-n(1+\omega_b)} \, . %
\label{rho vs a b}
\end{equation}

In this paper we assume $k_b=0$ for simplicity and study the
evolution of the scale factors $a(t)$ and $b(t)$ from Eqs.\
(\ref{G00 eq with extra dim})--(\ref{Gjj eq with extra dim}).
First of all, we consider a radiation-dominated universe with
\begin{eqnarray*}
\bar{p}_a &=& \frac{1}{3} \bar{\rho} \, ,\\
\bar{p}_b &=& 0 \, .
\end{eqnarray*}
For such a universe, we can read off, from Eqs.\ (\ref{G00 eq with
extra dim})--(\ref{Gjj eq with extra dim}), a solution with
constant $b$ (static extra dimensions), as shown in the following.
For constant $b$, Eqs.\ (\ref{G00 eq with extra dim})--(\ref{Gjj
eq with extra dim}) become
\begin{equation}
\left( \frac{\dot{a}}{a}\right)^2 + \frac{k_a}{a^2} = \frac{8 \pi
\bar{G}}{3} \bar{\rho} \, , %
\label{G00 eq with extra dim const b}
\end{equation}
\begin{equation}
2 \frac{\ddot{a}}{a} + \left( \frac{\dot{a}}{a} \right)^2 +
\frac{k_a}{a^2} = - \frac{8 \pi \bar{G}}{3} \bar{\rho} \, , %
\label{Gii eq with extra dim const b}
\end{equation}
\begin{equation}
\frac{\ddot{a}}{a} + \left( \frac{\dot{a}}{a} \right)^2 +
\frac{k_a}{a^2} = 0 \, .   %
\label{Gjj eq with extra dim const b}
\end{equation}
As a consistency check, we note that Eq.\ (\ref{Gjj eq with extra
dim const b}) can be derived from Eqs.\ (\ref{G00 eq with extra
dim const b}) and (\ref{Gii eq with extra dim const b}), the
latter being the equations that describe the evolution of a
four-dimensional radiation-dominated universe in the standard
cosmology. In addition, the constant $b$ solution is stable under
small perturbations of scale factors $a(t)$ and $b(t)$, a fact
that can be shown straightforwardly. In other words, we can
retrieve the ordinary evolution path of a radiation-dominated
universe within a higher-dimensional space-time with static extra
dimensions. On the other hand, for a matter-dominated universe
with $\bar{p}_a = \bar{p}_b = 0$, there is no solution
corresponding to constant $b$ (unless $\bar{\rho}=0$).

As a parenthetical remark, if we wish to obtain a solution with
the same evolution path as a four-dimensional matter-dominated
universe in the standard cosmology, that is, $b=$ const and $a(t)$
satisfies
\begin{equation}
\left( \frac{\dot{a}}{a} \right)^2 + \frac{k_a}{a^2} =\frac{8 \pi
\bar{G}}{3} \bar{\rho} \, ,
\end{equation}
\begin{equation}
2 \frac{\ddot{a}}{a} + \left( \frac{\dot{a}}{a} \right)^2 +
\frac{k_a}{a^2} = 0 \, ,
\end{equation}
the matter in the extra space needs to provide negative pressure
\[
\bar{p}_b = -\frac{1}{2} \bar{\rho} \; ,
\]
although it provides negligible pressure in the ordinary
three-space. It seems unusual to provide pressure in such a
strange manner. Nevertheless, it is in fact still possible that,
if the nonrelativistic ``particles'' in the ordinary three-space
are in fact extended objects (like strings) in the extra space,
they may provide pressure in this manner.

Assuming $k_a = k_b =0$ for simplicity and using the simple
equations of state $\bar{p}_a=\omega_a \bar{\rho}$ and
$\bar{p}_b=\omega_b \bar{\rho}$ with constant $\omega_a$ and
$\omega_b$, we obtain
\begin{equation}
\alpha (t) \equiv \left( 1-3\omega_a+2\omega_b \right)
\frac{\dot{a}}{a} - \left[ 1+(n-1)\omega_a-n\omega_b \right]
\frac{\dot{b}}{b} \propto \frac{1}{a^3 b^n} \propto
\frac{1}{V_{3+n}} \; , %
\label{alpha-volume relation}
\end{equation}
where $\alpha(t)$ in many cases characterizes roughly the
difference between the expansion rates of the ordinary three-space
and the extra space, and $V_{3+n}$ is the volume of the
$(3+n)$-dimensional space. It follows from Eq.\ (\ref{alpha-volume
relation}) that the difference $\alpha(t)$ will either decrease or
grow as the volume $V_{3+n}$ of the higher-dimensional world grows
or decreases.
For a radiation-dominated universe with
$\bar{p}_a=\frac{1}{3}\bar{\rho}$ and $\bar{p}_b=0$, we have
\begin{equation}
\alpha (t) = \frac{n+2}{3} \frac{\dot{b}}{b} \propto \frac{1}{a^3
b^n} \propto \frac{1}{V_{3+n}} \; . %
\label{alpha-volume for RD}
\end{equation}
Accordingly, if $V_{3+n}$ [the volume of overall (3+n)-dimensional
space] is growing, the expansion rate of the extra space will drop
and approach zero. We note that this feature also indicates the
stability of the constant-$b$ solution mentioned above in a
radiation-dominated universe. %
On the other hand, for a matter-dominated universe with
$\bar{p}_a=\bar{p}_b=0$ or for a more general case with
$\omega_a=\omega_b \neq 1/3$, we have
\begin{equation}
\alpha(t) \propto \frac{\dot{a}}{a} - \frac{\dot{b}}{b} \propto
\frac{1}{a^3 b^n} \propto \frac{1}{V_{3+n}} \; . %
\label{alpha-volume for MD}
\end{equation}
Accordingly, if the volume $V_{3+n}$ is growing, the expansion
rates of the ordinary three-space and the extra space will tend to
approach each other. If the volume $V_{3+n}$ is decreasing, on the
other hand, with one expanding space and one collapsing space,
then either $|\dot{a}/a|$ or $|\dot{b}/b|$ will become larger and
larger. We note that an increasing positive expansion rate
represents an accelerating expansion. Thus, the accelerating
expansion of our universe (the ordinary three-space) may be
generated along with the collapse of the extra space, as will be
discussed in the next section.


\section{Accelerating Expansion of the Ordinary Three-space Generated by
the Evolution of Extra Dimensions\label{accel 3-space}}


In this section, we consider a (1+3+n)-dimensional
matter-dominated universe and wish to explore, both analytically
and numerically, the possibility of generating the accelerating
expansion of the ordinary three-space via the the evolution of the
extra space. Assuming $k_a=k_b=0$ for simplicity and using the
equations of state, $\bar{p}_a=\bar{p}_b=0$, we can rearrange
Eqs.\ (\ref{Gii eq with extra dim}) and (\ref{Gjj eq with extra
dim}) to become
\begin{eqnarray}
(n+2) \frac{\ddot{a}}{a} + (2n+1) \left( \frac{\dot{a}}{a}
\right)^2 + n(n-1)\frac{\dot{a}}{a}\frac{\dot{b}}{b} -
\frac{n(n-1)}{2} \left( \frac{\dot{b}}{b} \right)^2 &=& 0 \; , \label{ddot-a} \\
(n+2) \frac{\ddot{b}}{b} - 3 \left( \frac{\dot{a}}{a} \right)^2 +
6 \frac{\dot{a}}{a}\frac{\dot{b}}{b} + \frac{(n-1)(n+4)}{2} \left(
\frac{\dot{b}}{b} \right)^2 &=& 0 \; , \label{ddot-b}
\end{eqnarray}
or, equivalently, with $u$ and $v$ defined as $u(t) \equiv
\dot{a}/a$ and $v(t) \equiv \dot{b}/b$,
\begin{eqnarray}
(n+2)\dot{u} + 3(n+1)u^2 + n(n-1)uv - \frac{n(n-1)}{2}v^2 &=& 0 \; , \label{dot-u} \\
(n+2)\dot{v} - 3u^2 + 6uv + \frac{n(n+5)}{2}v^2 &=& 0
\label{dot-v} \; .
\end{eqnarray}

From Eq.\ (\ref{ddot-a}), we can read off the condition for
$\ddot{a}/a > 0$ (accelerating ordinary three-space),
\begin{equation}
\frac{\dot{b}}{b} > \left[ 1 + \sqrt{\frac{(n+1)(n+2)}{n(n-1)}} \,
\right] \frac{\dot{a}}{a} \equiv J_{+} \left( \frac{\dot{a}}{a}
\right) \, , %
\label{ddot-a>0+}
\end{equation}
or
\begin{equation}
\frac{\dot{b}}{b} < \left[ 1 - \sqrt{\frac{(n+1)(n+2)}{n(n-1)}} \,
\right] \frac{\dot{a}}{a} \equiv J_{-} \left( \frac{\dot{a}}{a}
\right) \, , %
\label{ddot-a>0-}
\end{equation}
and also the condition for $\ddot{a}/a < 0$ (decelerating ordinary
three-space),
\begin{equation}
J_{-} \left( \frac{\dot{a}}{a} \right) < \frac{\dot{b}}{b} < J_{+}
\left( \frac{\dot{a}}{a} \right) .  %
\label{ddot-a<0}
\end{equation}
We can see that the fraction $\eta(t)$[$\equiv v(t)/u(t)$] is the
key quantity in the above conditions for swaying the acceleration
status of the ordinary three-space. Therefore it is essential to
investigate possible evolution paths of the fraction $\eta(t)$,
i.e., to see the behavior of $d\eta /dt$ in different regions of
$\eta$.

Using Eqs.\ (\ref{dot-u}) and (\ref{dot-v}), we obtain the
conditions
\begin{eqnarray}
\frac{d\eta}{dt} > 0 \; &\mbox{for}& \; \eta < K_{att}
\; \mbox{ or } \; K_{rep} < \eta < 1   \label{eta>0} \\
\frac{d\eta}{dt} < 0 \; &\mbox{for}& \; K_{att} < \eta < K_{rep}
\; \mbox{ or } \; \eta
> 1 \label{eta<0}
\end{eqnarray}
where $K_{att}$ and $K_{rep}$ are defined by
\begin{eqnarray}
K_{att} &=& - \frac{3+\sqrt{3(n+2)/n}}{n-1} \, , \\
K_{rep} &=& - \frac{3-\sqrt{3(n+2)/n}}{n-1} .
\end{eqnarray}
We note that the subscripts {\em att} and {\em rep} denote
``attractor'' and ``repeller'', respectively. Their meanings will
be described in the following discussion.

\begin{figure}[t]
\centerline{\psfig{figure=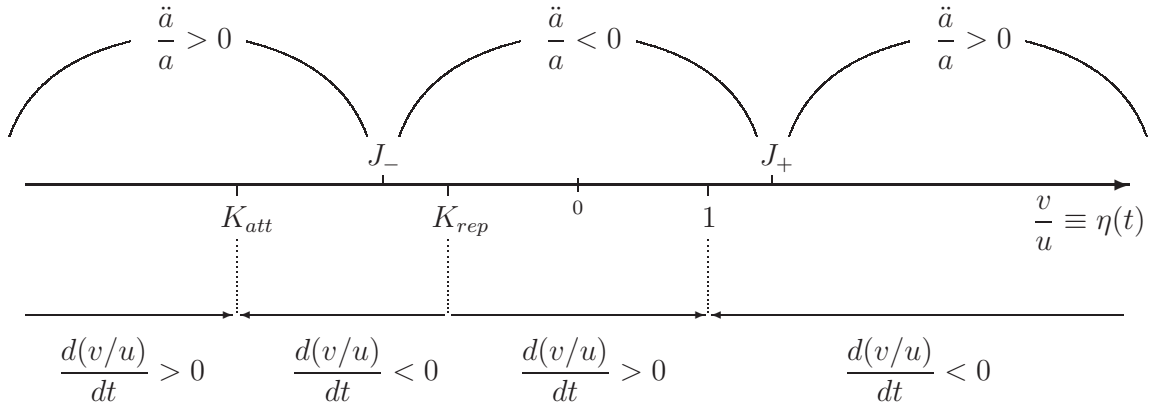,height=2.1in}} 
\caption{An illustration of (1) the situations for various
acceleration states of the ordinary three-space and (2) evolution
paths of the fraction $\eta = v/u$.} %
\label{accel-decel plot}
\end{figure}

The conditions in Eqs.\ (\ref{ddot-a>0+})--(\ref{eta<0}) can be
summarized in the ``flow diagram'' Fig.\ \ref{accel-decel plot}.
The upper part of Fig.\ \ref{accel-decel plot} illustrates the
conditions in Eqs.\ (\ref{ddot-a>0+})--(\ref{ddot-a<0}) for the
behavior of $\ddot{a}/a$, while the lower part illustrates the
conditions in Eqs.\ (\ref{eta>0}) and (\ref{eta<0}) for the
behavior of $d\eta /dt$. We note that $J_{+}$, $J_{-}$, $K_{att}$,
and $K_{rep}$ always obey the order
\begin{equation}
K_{att} < J_{-} < K_{rep} < 0 < 1 < J_{+}
\end{equation}
for $n \geq 2$. In Fig.\ \ref{accel-decel plot}, we can see that
there are two ``attractors'' at $\eta = K_{att}$ and at $\eta =
1$, and one ``repeller'' at $\eta = K_{rep}$. The
higher-dimensional world will approach the state at $\eta =
K_{att}$ for the class of initial conditions satisfying $\eta <
K_{rep}$, while it will approach $\eta = 1$ for the other
initial-condition class $\eta > K_{rep}$. Accordingly, the
higher-dimensional world possesses four kinds of evolution path
corresponding to different $\eta_0$ (the initial value of $\eta$):
\newcounter{temp}
\setcounter{temp}{\value{equation}}
\addtocounter{temp}{1}
\setcounter{equation}{0}
\renewcommand{\theequation}{\thetemp \alph{equation}}
\begin{eqnarray}
\eta_0 > J_{+} &,& \mbox{accelerate first and then decelerate,}
\label{path class 4} \\
K_{rep} < \eta_0 < J_{+} &,& \mbox{always
decelerate,} \label{path class 3} \\
J_{-} < \eta_0 < K_{rep} &,& \mbox{decelerate
first and then accelerate,} \label{decelerate-accelerate} \\
\eta_0 < J_{-} &,& \mbox{always accelerate.} \label{path class 1}
\end{eqnarray}
\addtocounter{temp}{1}
\renewcommand{\theequation}{\thetemp}
Consequently, we can see that the situation in Eq.\
(\ref{decelerate-accelerate}) may properly describe our universe.
In this situation, the universe is in the region $(J_{-} \, , \:
K_{rep})$ initially and the ordinary three-space expands in a
decelerating manner during the period when $\eta$ evolves within
this region, while the extra space always collapses. After some
time the ordinary three-space will start to accelerate as $\eta$
passes $J_{-}$ in the collapsing process of the extra space.
(Please see the {\em Note added}.)

\begin{figure}
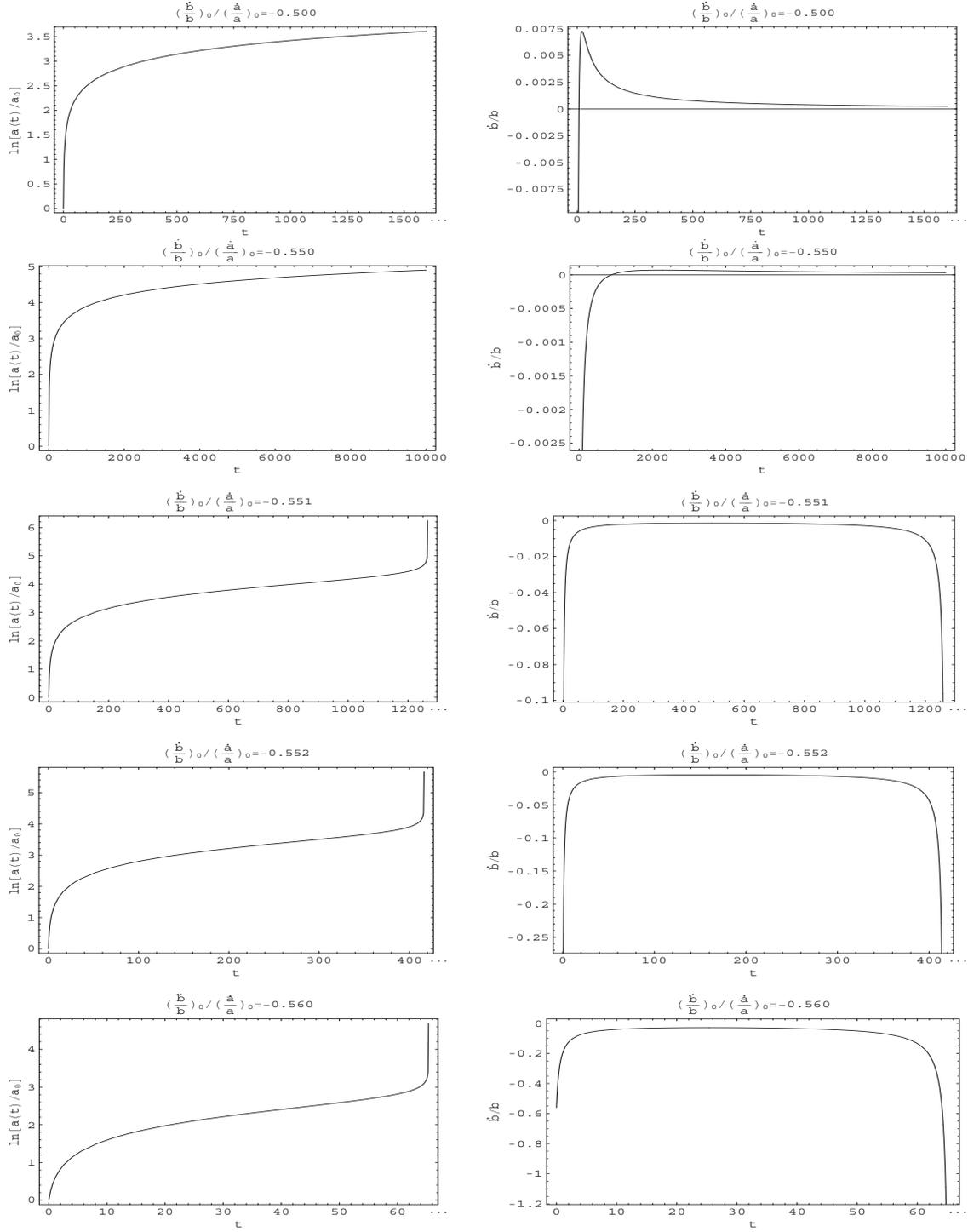

\begin{minipage}{2.8in}
\centerline{\psfig{figure=f1a.eps,width=2.7in,height=1.45in}} 
\end{minipage}
\hspace{0.4cm}
\begin{minipage}{2.8in}
\centerline{\psfig{figure=f1b.eps,width=2.95in,height=1.45in}}  
\end{minipage}
\\ \medskip
\begin{minipage}{2.8in}
\centerline{\psfig{figure=f2a.eps,width=2.7in,height=1.45in}} 
\end{minipage}
\hspace{0.4cm}
\begin{minipage}{2.8in}
\centerline{\psfig{figure=f2b.eps,width=2.95in,height=1.45in}}  
\end{minipage}
\\ \medskip
\begin{minipage}{2.8in}
\centerline{\psfig{figure=f3a.eps,width=2.7in,height=1.45in}} 
\end{minipage}
\hspace{0.4cm}
\begin{minipage}{2.8in}
\centerline{\psfig{figure=f3b.eps,width=2.95in,height=1.45in}}  
\end{minipage}
\\ \medskip
\begin{minipage}{2.8in}
\centerline{\psfig{figure=f4a.eps,width=2.7in,height=1.45in}} 
\end{minipage}
\hspace{0.4cm}
\begin{minipage}{2.8in}
\centerline{\psfig{figure=f4b.eps,width=2.95in,height=1.45in}}  
\end{minipage}
\\ \medskip
\begin{minipage}{2.8in}
\centerline{\psfig{figure=f5a.eps,width=2.7in,height=1.45in}} 
\end{minipage}
\hspace{0.4cm}
\begin{minipage}{2.8in}
\centerline{\psfig{figure=f5b.eps,width=2.95in,height=1.45in}}  
\end{minipage}
\caption{The evolution paths---plots of $\ln [a(t)/a_0]$ and
$\dot{b}/b$. The quantities $\dot{b}/b$ and $t$ are in units of
$(\dot{a}/a)_0$ and $(\dot{a}/a)_0^{-1}$, respectively, where the
subscript $0$ denotes the initial time $t_0 = t_{EQ}$ (which is set to be zero).} %
\label{ln-a & dot-b plot}
\end{figure}

\begin{figure}
\centerline{\psfig{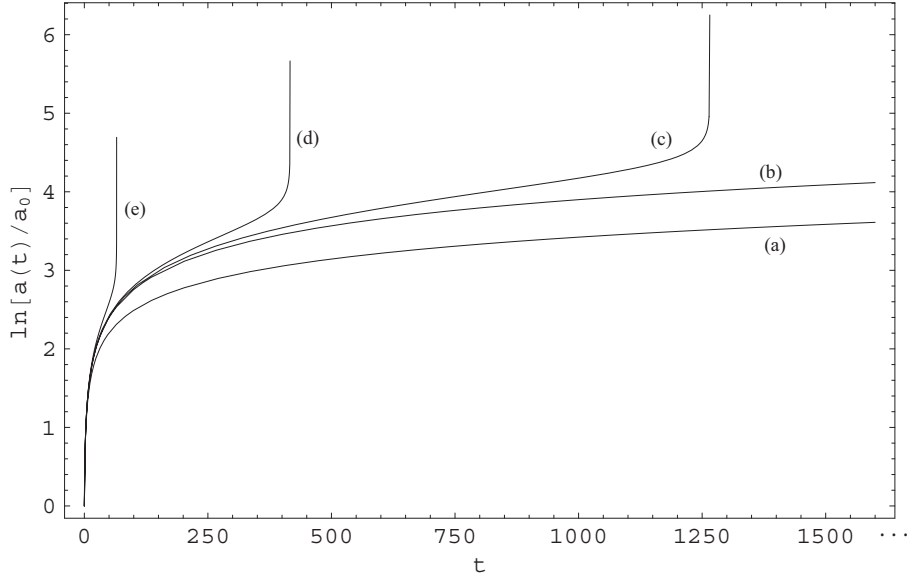}} 
\caption{The evolution paths---plot of $\ln [a(t)/a_0]$: (a)
$(\dot{b}/b)_0=-0.500(\dot{a}/a)_0 \,$, (b)
$(\dot{b}/b)_0=-0.550(\dot{a}/a)_0 \,$, (c)
$(\dot{b}/b)_0=-0.551(\dot{a}/a)_0 \,$, (d)
$(\dot{b}/b)_0=-0.552(\dot{a}/a)_0 \,$, (e)
$(\dot{b}/b)_0=-0.560(\dot{a}/a)_0 \,$. The time $t$ is in units
of $(\dot{a}/a)_0^{-1}$, where the subscript $0$
denotes the initial time $t_0 = t_{EQ}$ (which is set to be zero).} %
\label{ln-a-combine}
\end{figure}

For a concrete illustration, we analyze Eqs.\ (\ref{ddot-a}) and
(\ref{ddot-b}) numerically for the case of $n=2$. We consider the
initial time to be the moment when the nonrelativistic matter
starts to take over and dominate the universe, which is usually
denoted by $t_{EQ}$ (where EQ means ``matter-radiation
equality''). The results are demonstrated in Fig.\ \ref{ln-a &
dot-b plot} and Fig.\ \ref{ln-a-combine}, in which five evolution
paths corresponding to various initial conditions are drawn. The
five initial conditions we employ are (a)
$(\dot{b}/b)_0=-0.500(\dot{a}/a)_0$, (b)
$(\dot{b}/b)_0=-0.550(\dot{a}/a)_0$, (c)
$(\dot{b}/b)_0=-0.551(\dot{a}/a)_0$, (d)
$(\dot{b}/b)_0=-0.552(\dot{a}/a)_0$, and (e)
$(\dot{b}/b)_0=-0.560(\dot{a}/a)_0$. The plots in the left column
of Fig.\ \ref{ln-a & dot-b plot} sketch $\ln [a(t)/a_0]$, and the
plots in the right column sketch $\dot{b}(t)/b(t)$. Figure
\ref{ln-a-combine} combines the plots in the left column of Fig.\
\ref{ln-a & dot-b plot} for comparison. The value of $\dot{b}/b$
and the time $t$ are in units of $(\dot{a}/a)_0$ and
$(\dot{a}/a)_0^{-1}$ (the Hubble time at $t_0 = t_{EQ}$, which is
set to be zero here), respectively. We can see that the evolution
path is very sensitive to the initial value of $\eta$ and, more
importantly, there exists a critical value $\eta_{cr}$, which is
indeed the parameter $K_{rep}$ and is about $-0.55051$ for $n=2$,
such that the ordinary three-space will either expand in a
decelerating manner for $\eta_0 > \eta_{cr}$ or accelerate
eventually for $\eta_0 < \eta_{cr}$.


To sum up, the analysis of the matter-dominated case with
$k_a=k_b=0$ leads to Fig.\ \ref{accel-decel plot}. Four classes of
evolution path, as described in Eqs.\ (\ref{path class
4})--(\ref{path class 1}), can be read off from Fig.\
\ref{accel-decel plot}. The situation that may appropriately
describe our universe is the one suggested by Eq.\
(\ref{decelerate-accelerate}), in which the ordinary three-space
decelerates first and then accelerates along with the collapse of
the extra space. Figures \ref{ln-a & dot-b plot} and
\ref{ln-a-combine} show five evolution path corresponding to five
different initial conditions, where paths (a) and (b) illustrate
the situation in Eq.\ (\ref{path class 3}) and paths (c)--(e)
illustrate the situation in Eq.\ (\ref{decelerate-accelerate}).
There exists a critical value $\eta_{cr}$, which is exactly the
parameter $K_{rep}$, dividing these two classes of evolution
paths. For the case with $\eta_0 < \eta_{cr}$ corresponding to the
situation in Eq.\ (\ref{decelerate-accelerate}), the period before
the acceleration starts is sensitive to the the initial value of
$\eta$, $\eta_0$. If we require the expansion of our universe to
decelerate for a long enough period before the acceleration starts
in order to conserve the concordance between observations and
theories regarding the early universe, the initial value of $\eta$
has to be chosen in a delicate way such that it is close enough to
$\eta_{cr}$. Therefore this scenario so far has a fine-tuning
problem.


\section{Discussion and summary}
We have investigated the scenario of producing the accelerating
expansion of the present universe via evolving small extra
dimensions. For a radiation-dominated universe, such as our early
universe, we obtain a stable solution with static extra
dimensions. Accordingly, the existence of extra dimensions may
have no significant influence on the evolution of the ordinary
three-space. This is a good feature which we need for preserving
the concordance between observations and current theories
regarding the early (radiation-dominated) universe, especially for
primordial nucleosynthesis. On the contrary, such a solution with
static extra dimensions does not exist for the present
matter-dominated universe.

The features of the evolution can also be read off from Eq.\
(\ref{alpha-volume relation}), or Eqs.\ (\ref{alpha-volume for
RD}) and (\ref{alpha-volume for MD}), which are derived from Eq.\
(\ref{alpha-volume relation}). Equation (\ref{alpha-volume for
RD}) shows the decreasing expansion rate of the extra space along
with the increase of the $(3+n)$-dimensional volume $V_{3+n}$.
This implies the stability of the solution with static extra
dimensions in the radiation-dominated universe for the case of
$k_a=k_b=0$ as already mentioned above. On the other hand, Eq.\
(\ref{alpha-volume for MD}) shows two possible evolution patterns
of the matter-dominated universe: (i) The expansion rates of the
ordinary three-space and the extra space tend to catch up with
each other along with the increase of the (3+n)-dimensional volume
$V_{3+n}$. (ii) One of these two expansion rates is positive and
increasing, while the other is negative and decreasing, along with
the decrease of the (3+n)-dimensional volume $V_{3+n}$. We note
that an increasing positive expansion rate represents an
accelerating expansion.

A quantitative analysis of the matter-dominated case with
$k_a=k_b=0$ leads to Fig.\ \ref{accel-decel plot}, which indicates
four classes of evolution path. A universe that decelerates first
and then accelerates is included in one of them. Therefore the
accelerating expansion of the present universe may be
appropriately described in this scenario. In addition, the case
with two extra dimensions is analyzed in detail. The five
resultant evolution paths we draw demonstrate the existence of a
critical value for the initial condition $\eta_0$, which divides
two classes of path: the one in which the universe decelerates
first and then accelerates and the other in which the universe
always decelerates. We note that the critical value $\eta_{cr}$ is
exactly the parameter $K_{rep}$, a ``repeller'' in the flow
diagram. However, the existence of the critical value (or the
``repeller'') also implies a fine-tuning problem, i.e., the
initial value of $\eta$ has to be chosen delicately so that it is
close enough to the critical value $\eta_{cr}$ in order to possess
a long enough decelerating epoch followed by an accelerating
epoch.

The existence of extra dimensions is a general feature in theories
beyond the standard model in particle physics. It may manifest
itself as a source of energy in the ordinary three-space, such as
``effective'' dark energy or even ``effective'' dark matter. The
geometrical structure and the evolution pattern of extra
dimensions therefore may play an important role in cosmology. In
this work we study a simple scenario of extra dimensions that is
subject to a fine-tuning problem. Nevertheless, other scenarios
with richer structures, such as those in \cite{Arkani-Hamed} and
\cite{Randall&Sundrum}, may also provide suitable evolution
patterns and are worthy of being further investigated.

{\em Note added}. For the sake of simplicity, we have in this
paper considered only the case with $k_a=k_b=0$, i.e., both our
ordinary three-space and the extra space are flat. The general
situations with nonzero $k_a$ or $k_b$ clearly may offer many
interesting possibilities and are currently under serious
investigation.

\section*{Acknowledgements}
One of us (Je-An Gu) wishes to thank Professor W.~F.~Kao for
helpful discussions. This work was supported in part by the
National Science Council, Taiwan, R.O.C.\ (NSC 90-2112-M-002-028)
and by the CosPA project of the Ministry of Education (MOE
89-N-FA01-1-4-3).

\end{document}